\documentclass[preprint]{elsarticle}

\usepackage{fullpage}
\usepackage[english]{babel}
\usepackage[utf8]{inputenc}
\usepackage{amsmath, amssymb, bm}
\usepackage{graphicx}
\usepackage{subcaption}
\usepackage[colorinlistoftodos]{todonotes}
\usepackage{indentfirst}
\usepackage{hyperref}

\newcommand{\hiddenpower}[2] { \ifnum \numexpr#2=1 #1 \else #1^#2 \fi }
\numberwithin{equation}{section}

\newcommand{\dd}{\text{d}}
\newcommand{\pd}{\partial}

\usepackage{xparse}
\ExplSyntaxOn
\newcounter{diff_order}
\newcounter{diff_power}

\newcommand{\rawdiff}[3]
{
	\setcounter{diff_order}{0}
	\clist_map_inline:nn{#3}{\stepcounter{diff_order}}
	
	\frac{\hiddenpower{#1}{\thediff_order} #2}
	{
		\def\old_var{DefaultValue}
		\setcounter{diff_power}{0}
		
		\clist_map_inline:nn{#3}
		{
			\def\new_var{##1}
			\ifnum \thediff_power=0
				\stepcounter{diff_power}
			\else
				\tl_if_eq:NNTF \new_var \old_var
				{\stepcounter{diff_power}}
				{
					#1 \hiddenpower{\old_var}{\thediff_power}
					\setcounter{diff_power}{1}
				}
			\fi

			\def\old_var{##1}
		}
		
		#1 \hiddenpower{\old_var}{\thediff_power}
	}
}

\ExplSyntaxOff

\newcommand{\lb}{\left(}
\newcommand{\rb}{\right)}

\renewcommand{\ln}[1]{\text{ln} \lb #1 \rb}
\newcommand{\e}[1]{\text{e}^{#1}}

\newcommand{\tr}[1]{\text{tr}\left\lbrace #1 \right\rbrace}
\newcommand{\ptr}[2]{\text{tr}_{#1}\left\lbrace #2 \right\rbrace}

\newcommand{\set}[1]{\ensuremath{\left\lbrace #1 \right\rbrace}}

\newcommand{\nbym}[2]{\ensuremath{#1{\times}#2}}
\newcommand{\nbyn}[1]{\nbym{#1}{#1}}

\newcommand{\sln}[1]{\ensuremath{\mathfrak{sl}_{#1}}}

\title{A Dual Construction of the Isotropic Landau-Lifshitz Model}

\author[1]{Iain Findlay}
\ead{iaf1@hw.ac.uk}

\address[1]{School of Mathematical and Computer Sciences, \\ Heriot-Watt University, Edinburgh, EH14 4AS, United Kingdom}

\begin{document}

\begin{abstract}
	By interchanging the roles of the space and time coordinates, we describe a dual construction of the isotropic Landau-Lifshitz model, providing equal-space Poisson brackets and dual Hamiltonians conserved with respect to space-evolution. This construction is built in the Lax/zero-curvature formalism, where the duality between the space and time dependencies is evident.
\end{abstract}


\maketitle

\section{Introduction}
\label{sec:Intro}

The idea of considering 1+1 dimensional integrable models in terms of their ``space-evolution", as governed by some equal-space Poisson brackets found by interchanging the roles of the space and time coordinates was systematically introduced in \cite{ref:acdk_dual}, following the suggestion in \cite{ref:ck_dual} for the purposes of identifying integrable defects (that lie in the spatial axis) with Darboux-B\"acklund transformations. This concept was applied rigorously to the Lax/zero-curvature construction \cite{ref:l_lax, ref:akns_lax} of the non-linear Schr\"odinger (NLS) model in \cite{ref:acdk_dual}, and then later proven for the general NLS hierarchy in \cite{ref:ac_dual}.

In this paper, we apply this equal-space construction to the isotropic Landau-Lifshitz model \cite{ref:l_hm, ref:t_hm}, which is also known as the continuous classical Heisenberg magnet (HM) model:
\begin{equation}
	\pd_t \vec{S} = \frac{\text{i}}{c^2} \vec{S} \times (\pd_x^2 \vec{S}), \label{eq:EoMs_vec}
\end{equation}

\noindent which depends on the vector $ \vec{S} = (S_x, S_y, S_z)^T $. These fields will also be written in the combinations $ S_{\pm} = S_x \pm \text{i} S_y $, which satisfy the \sln{2} exchange relations:
\begin{equation}
	\lbrace S_{\pm}(x), S_z(y) \rbrace = \pm S_{\pm} \,\delta(x - y), \qquad\qquad\qquad \lbrace S_{+}(x), S_{-}(y) \rbrace = -2S_z \,\delta(x - y). \label{eq:x_PBs}
\end{equation}

\noindent These Poisson brackets are found through the $ r $-matrix construction \cite{ref:s_linalg}. The HM model has the same underlying $ r $-matrix as the NLS model, namely the Yangian $ r $-matrix \eqref{eq:rMat}, so hence it arises as a natural next step in the development of this dual approach.

Because this equal-space picture follows in parallel to the usual method for building conserved quantities and higher systems (see \cite{ref:ft_book}), we also introduce reflective time-like boundary conditions \cite{ref:dfs_nls} to the HM model by following an equivalent procedure to the development of reflective space-like boundary conditions, \cite{ref:s_qbcs, ref:s_bcs}, which have been applied to the isotropic Landau-Lifshitz equation in \cite{ref:dk_ll}.

The HM model is also of recent practical interest as a simple model of 1 dimensional ferromagnetism (due to being the continuum limit of the classical analogue of the quantum XXX spin chain, see \cite{ref:ft_book, ref:ads_hmlim, ref:f_book} for details), \cite{ref:dlsvv_hm, ref:msp+_magnets, ref:ls_magnets}. This paper therefore sheds new light on this model by approaching it from a time-like perspective, analogous to the standard description in terms of time-evolution.

\-

The paper is laid out as follows: The remainder of Section \ref{sec:Intro} defines the basic terms that we will be using throughout. Then, in Section \ref{sec:Usual} we describe the standard (equal-time) construction of the hierarchy of conserved quantities and their associated Lax pairs and integrable systems, applying these to the HM model for later comparison. This section starts by constructing the Poisson brackets between the fields in Subsection \ref{ssec:x_PBs}, before building the hierarchy of conserved quantities that guarantee the integrability of the HM model. This is done for both closed (periodic) boundary conditions in Subsection \ref{ssec:x_Closed} and open (reflective) boundary conditions in Subsection \ref{ssec:x_Open}. Subsection \ref{ssec:x_Open} recalls the results of \cite{ref:dk_ll}, except using notation that will be consistent with the sections that follow. Finally, we repeat these same steps for the dual (equal-space) construction of the HM model in Section \ref{sec:Dual}, with the dual Poisson structure constructed in Subsection \ref{ssec:t_PBs}, and the hierarchies of dual Hamiltonians (and the corresponding Lax pairs) for both closed and open boundary conditions are constructed in Subsections \ref{ssec:t_Closed} and \ref{ssec:t_Open}, respectively.

\-

\subsection{Preliminaries}
\label{ssec:Prelims}

In terms of the fields $ S_{\pm} $ and $ S_z $, the equations of motion \eqref{eq:EoMs_vec} become:
\begin{equation}
	\pd_t S_{\pm} = \pm \frac{1}{c^2} \big( S_{\pm} (\pd_x^2 S_z) - (\pd_x^2 S_{\pm}) S_z \big), \qquad\qquad\qquad \pd_t S_z = \frac{1}{2c^2} \big( (\pd_x^2 S_{+}) S_{-} - S_{+} (\pd_x^2 S_{-}) \big). \label{eq:EoMs}
\end{equation}

\noindent When referencing the three fields $ S_{\pm} $ and $ S_z $, we will use the subscript $ \sigma \in \set{+, -, z} $ to collectively refer to them as $ S_{\sigma} $. We will also use $ \dot{S}_{\sigma} = \pd_{t_k} S_{\sigma} $ to denote the derivative of $ S_{\sigma} $ with respect to the appropriate time flow\footnote{These distinct time flows will arise from considering the tower of conserved quantities that define the system as integrable, and treating each of the quantities as the Hamiltonian for a distinct integrable system, describing the evolution of the fields along the associated time flow $ t_k $. When we consider the dual picture, we will likewise have a hierarchy of dual Hamiltonians that govern the space-evolution of the fields along a tower of space flows $ x_k $.} $ t_k $, and $ S_{\sigma}' = \pd_{x_k} S_{\sigma} $ for the derivative with respect to the contextually appropriate space flow. Where there is likely ambiguity however, we will explicitly use either $ \pd_{t_k} $ or $ \pd_{x_k} $.

It was shown in \cite{ref:t_hm} that the system of equations \eqref{eq:EoMs} appear as the compatibility condition of the auxiliary linear problem:
\begin{equation}
	\Psi' \equiv \pd_x \Psi = U \Psi, \qquad\qquad\qquad \dot{\Psi} \equiv \pd_t \Psi = V \Psi, \label{eq:ALP}
\end{equation}

\noindent where $ \Psi $ is an arbitrary vector field, and the \nbyn{2} matrices $ U $ and $ V $, depending on the fields $ S_{\sigma} $ as well as some free complex parameter $ \lambda $, comprise the Lax pair \cite{ref:l_lax, ref:akns_lax} of the system, given by:
\begin{equation}
	U = \frac{1}{2\lambda} S, \qquad\qquad\qquad V = \frac{1}{2\lambda^2} S - \frac{1}{2c^2 \lambda} S' S, \label{eq:Lax}
\end{equation}

\noindent where:
\begin{equation}
	S = \lb \begin{matrix}
		S_z & S_{-} \\
		S_{+} & -S_z
	\end{matrix} \rb. \nonumber
\end{equation}

\noindent Cross-differentiating the auxiliary linear problem gives rise to the following compatibility condition (called the zero-curvature condition) between the matrices of the Lax pair:
\begin{equation}
	0 = \dot{U} - V' + [U, V], \label{eq:ZCC}
\end{equation}

\noindent such that when the matrices $ U $ and $ V $ are inserted into this, and the resulting equations are split about powers of $ \lambda $, the equations of motion, \eqref{eq:EoMs}, are returned.

\-

\section{The Standard Picture}
\label{sec:Usual}

\subsection{Poisson Brackets}
\label{ssec:x_PBs}

Before we introduce the dual picture for \eqref{eq:EoMs} we first recap the method for constructing the hierarchy of integrable equations and their Hamiltonians. The core objects in this construction are the spatial component of the Lax pair, $ U $, and an associated $ r $-matrix that satisfies the classical Yang-Baxter equation \cite{ref:sts_vgen}:
\begin{equation}
	0 = [r_{ab}(\lambda - \mu), r_{ac}(\lambda)] + [r_{ab}(\lambda - \mu), r_{bc}(\mu)] + [r_{ac}(\lambda), r_{bc}(\mu)], \label{eq:cYBE}
\end{equation}

\noindent where $ \lambda, \mu \in \mathbb{C} $ are some free parameters and the subscripts denote which vector spaces the matrices act on (e.g. $ r_{ab} = r \otimes \mathbb{I} $ and $ r_{bc} = \mathbb{I} \otimes r $, with $ r : V \otimes V \to V \otimes V $, so that the whole equation acts on $ V_a \otimes V_b \otimes V_c $, where the subscripts attached to the vector spaces are merely used to denote which index corresponds to them, e.g. $ r_{ab} $ would act only on the first two). For the HM model, the relevant solution is:
\begin{equation}
	r(\lambda) = \frac{1}{2\lambda} \lb \begin{matrix} 1 & 0 & 0 & 0 \\ 0 & 0 & 1 & 0 \\ 0 & 1 & 0 & 0 \\ 0 & 0 & 0 & 1 \end{matrix} \rb. \label{eq:rMat}
\end{equation}

This $ r $-matrix is connected to the $ U $-matrix and the equations of motion for the system \eqref{eq:EoMs} through the linear algebraic relation\footnote{The subscript $ {}_S $ is used here and in what follows to denote that we are building this system out of the \underline{S}patial component of the Lax pair ($ U $). This will be important later when we construct the dual model out of the \underline{T}emporal component of the Lax pair ($ V $), where we will use a $ {}_T $ subscript.} \cite{ref:s_linalg}:
\begin{equation}
	\lbrace U_a(x, \lambda), U_b(y, \mu) \rbrace^{\-}_S = [r_{ab}(\lambda - \mu), U_a(x, \lambda) + U_b(y, \mu)] \,\delta(x - y), \label{eq:x_LinAlg}
\end{equation}

\noindent which provides an ultra-local Poisson bracket between the fields. Inserting the $ U $-matrix \eqref{eq:Lax} and $ r $-matrix \eqref{eq:rMat} into this relation returns the \sln{2} exchange relations, \eqref{eq:x_PBs}. From these Poisson brackets we can read off a Casimir element that restricts the vector $ \vec{S} $ to the surface of the sphere of radius $ c $, where we have labelled the Casimir $ c^2 $:
\begin{equation}
	c^2 = S_z^2 + S_{+} S_{-} = S_x^2 + S_y^2 + S_z^2. \label{eq:Casimir}
\end{equation}

\-

\subsection{Periodic Boundary Conditions}
\label{ssec:x_Closed}

In order to find conserved quantities that commute with respect to this Poisson bracket, we start by considering the (spatial) transport matrix, which is a path-ordered exponential solution to the spatial component of the auxiliary linear problem \eqref{eq:ALP} in place of $ \Psi $:
\begin{equation}
	T_S(x, y; \lambda) = \text{P} \exp{\int_{y}^{x} U(\xi) \dd \xi}. \label{eq:x_Monom}
\end{equation}

\noindent For a periodic system on the interval $ [-L, L] $, i.e. where $ S_{\sigma}(L) = S_{\sigma}(-L) $, the full monodromy matrix is $ T_S(\lambda) = T_S(L, -L; \lambda) $. Due to the $ U $-matrices satisfying the linear algebraic relation, \eqref{eq:x_LinAlg}, the monodromy matrix can be seen to satisfy a quadratic algebraic relation \cite{ref:stf_frt, ref:frt_frt}:
\begin{equation}
	\lbrace T_{S, a}(\lambda), T_{S, b}(\mu) \rbrace^{\-}_S = [r_{ab}(\lambda - \mu), T_{S, a}(\lambda) T_{S, b}(\mu)]. \label{eq:x_QuadAlg}
\end{equation}

\noindent Consequently, if we define a new object, called the transfer matrix $ \mathfrak{t}_S(\lambda) $, as the trace of the monodromy matrix:
\begin{equation}
	\mathfrak{t}_S(\lambda) = \tr{T_S(\lambda)}, \label{eq:x_transfer}
\end{equation}

\noindent then this can be shown to Poisson commute with itself for different values of the spectral parameter $ \lambda $. Because of this, if we expand $ \mathfrak{t}_S $ as a formal power series in $ \lambda $, $ \mathfrak{t}_S = \sum_k \lambda^k \mathfrak{t}_S^{(k)} $, then these coefficients commute:
\begin{equation}
	\lbrace \mathfrak{t}_S^{(k)}, \mathfrak{t}_S^{(j)} \rbrace^{\-}_S = 0. \label{eq:x_HamsComm}
\end{equation}

\noindent As such, the terms in this expansion $ \mathfrak{t}_S^{(k)} $ can be seen as ``Hamiltonians" governing the evolution of the system along distinct time flows $ t_k $. Further to this, the evolution along each time flow $ t_k $ will be integrable \textit{\`a la} Liouville, as the $ \mathfrak{t}_S^{(j)} $ with $ j \neq k $ will provide the infinite tower of conserved quantities.

Unfortunately, the ``Hamiltonians" generated in this manner will be non-local. To circumvent this, we will consider the coefficients in the expansion of the logarithm of this, $ \mathcal{G}_S(\lambda) = \ln{\mathfrak{t}_S(\lambda)} $. The logarithm is chosen as it acts to remove the non-locality introduced by the exponential in \eqref{eq:x_Monom} and in the diagonalisation below, \eqref{eq:x_WZW}.

\-

The task is therefore to find the expansion of $ \mathfrak{t}_S(\lambda) $ in some limit of $ \lambda $. For the Lax pair \eqref{eq:Lax} the appropriate limit is $ \lambda \to 0^{+} $. In order to avoid evaluating the path-ordered exponential, we consider a diagonalisation of the transport matrix \cite{ref:ft_book}:
\begin{equation}
	T_S(x, y; \lambda) = \big( \mathbb{I} + W_S(x; \lambda) \big) \e{Z_S(x, y; \lambda)} \big( \mathbb{I} + W_S(y; \lambda) \big)^{-1}, \label{eq:x_WZW}
\end{equation}

\noindent where $ W_S $ and $ Z_S $ are wholly anti-diagonal and diagonal matrices, respectively. If we insert this diagonalisation into the spatial half of the auxiliary linear problem, the diagonal and anti-diagonal components can be separated into two relations:
\begin{equation}
\begin{aligned}
	0 &= W_S' + [W_S, U_D] + W_S U_A W_S - U_A, \\
	Z_S' &= U_D + U_A W_S,
\end{aligned} \label{eq:x_WZ}
\end{equation}

\noindent where $ U_D $ and $ U_A $ are the diagonal and anti-diagonal components of the $ U $-matrix, respectively. If we expand $ W_S $ and $ Z_S $ in powers of $ \lambda $, with coefficients $ W_S^{(k)} $ and $ Z_S^{(k)} $ \cite{ref:ft_book}:
\begin{equation}
	W_S(\lambda) = \sum_{k = 0}^{\infty} \lambda^k W_S^{(k)}, \qquad\qquad\qquad Z_S(\lambda) = \sum_{k = -1}^{\infty} \lambda^k Z_S^{(k)}, \nonumber
\end{equation}

\noindent we can split \eqref{eq:x_WZ} into a series of recurrence relations (making use of how $ U $ only depends on $ \lambda^{-1} $):
\begin{equation}
\begin{gathered}
	0 = [W_S^{(0)}, U_D] + W_S^{(0)} U_A W_S^{(0)} - U_A, \qquad\qquad 0 = (W_S^{(k)})' + [W_S^{(k + 1)}, U_D] + \sum_{j = 0}^{k + 1} W_S^{(k + 1 - j)} U_A W_S^{(j)}, \\
	(Z_S^{(-1)})' = U_D + U_A W_S^{(0)}, \qquad\qquad (Z_S^{(k)})' = U_A W_S^{(k + 1)},
\end{gathered} \nonumber
\end{equation}

\noindent which we can recursively solve to find ever higher coefficients in the series expansions of $ W_S $ and $ Z_S $. The first few terms in the $ Z_S $-series are:
\begin{equation}
\begin{aligned}
	Z_S^{(-1)} &= cL \lb \begin{matrix}
		1 & 0 \\
		0 & -1
	\end{matrix} \rb, \\
	Z_S^{(0)} &= \frac{1}{4c} \int_{-L}^{L} \frac{S_{+} S_{-}' - S_{+}' S_{-}}{c + S_z} \lb \begin{matrix}
		1 & 0 \\
		0 & -1
	\end{matrix} \rb \dd x, \\
	Z_S^{(1)} &= \frac{-1}{4c^3} \int_{-L}^{L} \big( S_{+}' S_{-}' + (S_z')^2 \big) \lb \begin{matrix}
		1 & 0 \\
		0 & -1
	\end{matrix} \rb \dd x.
\end{aligned} \label{eq:x_Zs}
\end{equation}

\-

The reason for doing this is that if we insert the decomposition into the definition of the transfer matrix, \eqref{eq:x_transfer}, the explicit $ W $ dependence cancels out, leaving:
\begin{equation}
	\mathfrak{t}_S(\lambda) = \tr{\e{Z_S(\lambda)}} = \e{Z_{11, S}(\lambda)} + \e{Z_{22, S}(\lambda)}. \nonumber
\end{equation}

\noindent We are actually instead interested in the expansion of $ \mathcal{G}_S = \ln{\mathfrak{t}_S} $, which is then:
\begin{equation}
	\mathcal{G}_S(\lambda) = \ln{\e{\lambda^{-1} Z_{11, S}^{(-1)} + Z_{11, S}^{(0)} + \lambda Z_{11, S}^{(1)} + ...} + \e{\lambda^{-1} Z_{22, S}^{(-1)} + Z_{22, S}^{(0)} + \lambda Z_{22, S}^{(1)} + ...}}. \nonumber
\end{equation}

\noindent As the leading order terms in each of the exponents are $ cL \lambda^{-1} $ and $ -cL \lambda^{-1} $, and we are considering the limit as $ \lambda \to 0^{+} $, the first exponential will be of the form $ \e{cL \lambda^{-1}} $, so will dominate over the second exponential, which will be of the form $ \e{-cL \lambda^{-1}} $, which decays exponentially in the limit $ \lambda \to 0^{+} $. The expansion of $ \mathcal{G}_S(\lambda) $ is therefore simply:
\begin{equation}
	\mathcal{G}_S(\lambda) = \lambda^{-1} Z_{11, S}^{(-1)} + Z_{11, S}^{(0)} + \lambda Z_{11, S}^{(1)} + .... \nonumber
\end{equation}

\noindent The first three conserved quantities appearing in this expansion can then be read from the $ Z $-series:
\begin{equation}
\begin{aligned}
	\mathcal{G}_S^{(-1)} &= cL, \\
	\mathcal{G}_S^{(0)} &= \frac{1}{4c} \int_{-L}^{L} \frac{S_{+} S_{-}' - S_{+}' S_{-}}{c + S_z} \dd x, \\
	\mathcal{G}_S^{(1)} &= \frac{-1}{4c^3} \int_{-L}^{L} \big( S_{+}' S_{-}' + (S_z')^2 \big) \dd x,
\end{aligned} \label{eq:x_Hams}
\end{equation}

\noindent the second and third of which can be recognised as the total momentum and Hamiltonian for the HM model, respectively (up to a factor of $ -2c $) \cite{ref:ft_book}:
\begin{equation}
	P_S = -2c \mathcal{G}_S^{(0)}, \qquad\qquad\qquad H_S = -2c \mathcal{G}_S^{(1)}. \label{eq:x_PH}
\end{equation}

\-

Each of the conserved quantities $ \mathcal{G}_S^{(k)} $ generated through the expansion of $ \mathcal{G}_S $ can be seen to describe the evolution of the system along a distinct time flow $ t_k $, so that the equations of motion for each of these systems would be given by:
\begin{equation}
	\pd_{t_k} S_{\sigma} = \lbrace \mathcal{G}_S^{(k)}, S_{\sigma} \rbrace^{\-}_S. \label{eq:TimeFlows}
\end{equation}

\noindent Consequently, each of these systems should have some associated Lax pair. As we use the $ U $-matrix to generate the conserved quantities we will be looking for a generator $ \mathbb{V} $ that produces the $ V $-matrices $ V^{(k)} $ associated to each time flow $ t_k $. We do so by first equating Hamilton's equation (as applied to $ U $) and the zero-curvature condition:
\begin{equation}
\begin{aligned}
	\mathbb{V}_b'(\lambda, \mu) - [U_b(\lambda), \mathbb{V}_b(\lambda, \mu)] &= \pd_{\bar{t}} U_b(\lambda) = \lbrace \ln{\ptr{a}{T_{S, a}(\mu)}}, U_b(\lambda) \rbrace_S \\
	&= \mathfrak{t}_S^{-1}(\mu) \ptr{a}{\lbrace T_{S, a}(\mu), U_b(\lambda) \rbrace_S}
\end{aligned} \nonumber
\end{equation}

\noindent where the $ \bar{t} $ is used to denote some master time flow and the vector space subscripts are introduced to distinguish the space being traced over (the $ a $ vector space). Using the algebraic relations \eqref{eq:x_LinAlg} and \eqref{eq:x_QuadAlg}, we can extract from this the generator of the $ V $-matrices associated to each time flow $ t_k $, \cite{ref:sts_vgen}:
\begin{equation}
	\mathbb{V}_b(x; \lambda, \mu) = \mathfrak{t}_S^{-1}(\mu)\, \ptr{a}{T_{S, a}(L, x; \mu) r_{ab}(\mu - \lambda) T_{S, a}(x, -L; \mu)}, \label{eq:x_STS}
\end{equation}

\noindent such that the $ V $-matrix associated to the $ t_k $ time flow appears as the coefficient of $ \mu^k $ in the series expansion of this about $ \mu $. Using the diagonalisation of the monodromy matrix, the limit $ \mu \to 0^{+} $ of the exponential of $ Z_S(\mu) $, and the cyclic properties of the trace, this can be simplified to:
\begin{equation}
	\mathbb{V}_b(x; \lambda, \mu) = \ptr{a}{r_{ab}(\mu - \lambda) \big( \mathbb{I} + W_{S, a}(x; \mu) \big) e_{11, a} \big( \mathbb{I} + W_{S, a}(x; \mu) \big)^{-1}}, \nonumber
\end{equation}

\noindent where $ e_{ij} $ is the \nbyn{2} matrix that obeys $ (e_{ij})_{kl} = \delta_{ik} \delta_{jl} $. Finally, as the chosen $ r $-matrix satisfies $ r_{ab} M_a = M_b r_{ab} $ for any \nbyn{2} matrix $ M $, this can be simplified further to lie solely in the $ b $ vector space (so that we may drop the subscripts):
\begin{equation}
	\mathbb{V}(x; \lambda, \mu) = \frac{1}{\mu - \lambda} \big( \mathbb{I} + W_S(x; \mu) \big) e_{11} \big( \mathbb{I} + W_S(x; \mu) \big)^{-1}. \label{eq:x_STS_final}
\end{equation}

\noindent If we expand this about powers of $ \mu $ in the limit as $ \mu \to 0^{+} $, the first three terms are:
\begin{equation}
\begin{aligned}
	\mathbb{V}^{(0)} &= \frac{-1}{4\lambda} \mathbb{I} - \frac{1}{4c\lambda} S, \\
	\mathbb{V}^{(1)} &= \frac{-1}{4\lambda^2} \mathbb{I} - \frac{1}{4c\lambda^2} S + \frac{1}{4c^3 \lambda} S'S, \\
	\mathbb{V}^{(2)} &= \frac{-1}{4\lambda^3} \mathbb{I} - \frac{1}{4c\lambda^3} S + \frac{1}{4c^3 \lambda^2} S'S - \frac{1}{4c^3 \lambda} S'' - \frac{3}{8c^5 \lambda} (S')^2 S.
\end{aligned} \label{eq:VMats}
\end{equation}

\noindent After removing the overall commuting constant factors and scaling by $ -2c $, the second of these can be identified as the $ V $-matrix in the Lax pair \eqref{eq:Lax}:
\begin{equation}
	V = -2c (\mathbb{V}^{(1)} + \frac{1}{4\lambda^2} \mathbb{I}). \nonumber
\end{equation}

\-

It is the identification of $ U $ with $ \mathbb{V}^{(0)} $ up to some constant factors, that suggests the introduction of a dual picture for this model, with the roles of time and space switched. Before we investigate this though, we briefly discuss how to adapt this construction to account for non-periodic boundary conditions.

\-

\subsection{Open Boundary Conditions}
\label{ssec:x_Open}

In order to study systems with open boundary conditions, we need to introduce some $ K_{\pm} $-matrices that are associated to the $ \pm L $ boundaries, and have a dependence on the spectral parameter and some additonal constants. In order for them to be used in generating conserved quantities, we require that they satisfy the classical analogue of the (non-dynamical) quantum reflection equation \cite{ref:s_bcs}:
\begin{equation}
	0 = [r_{ab}(\lambda - \mu), K_{\pm, a}(\lambda) K_{\pm, b}(\mu)] + K_{\pm, a}(\lambda) r_{ab}(\lambda + \mu) K_{\pm, b}(\mu) - K_{\pm, b}(\mu) r_{ab}(\lambda + \mu) K_{\pm, a}(\lambda). \label{eq:cRA}
\end{equation}

\noindent For the $ r $-matrix \eqref{eq:rMat}, the most general choice of $ K_{\pm} $-matrix (up to some rescaling and gauge transformations) is \cite{ref:dg_kmat}:
\begin{equation}
	K_{\pm}(\lambda) = \alpha_{\pm} \mathbb{I} + \lambda \lb \begin{matrix}
		\delta_{\pm} & \beta_{\pm} \\
		\gamma_{\pm} & -\delta_{\pm}
	\end{matrix} \rb, \label{eq:KMat}
\end{equation}

\noindent where $ \alpha_{\pm} $, $ \beta_{\pm} $, $ \gamma_{\pm} $, and $ \delta_{\pm} $ are some constants that describe the boundary conditions being considered\footnote{The reflection equation satisfied by the $ K_{+} $- and $ K_{-} $-matrices actually differ by a minus sign in the spectral parameter, but we absorb this factor into the $ \beta_{+} $, $ \gamma_{+} $, and $ \delta_{+} $ to keep the forms of the matrices the same.}. If these are given a time dependence, then these would be dynamical boundary conditions. For this paper, however, we consider only the non-dynamical case where they have no time dependence (and when we move on to discuss time-like boundary conditions, we shall assume that the equivalent constants have no space dependence). These $ K_{\pm} $-matrices are introduced into the transfer matrix $ \mathfrak{t}_S $ as \cite{ref:s_qbcs, ref:s_bcs}:
\begin{equation}
	\bar{\mathfrak{t}}_S(\lambda) = \tr{K_{+}(\lambda) T_S(L, -L; \lambda) K_{-}(\lambda) T_S^{-1}(L, -L; -\lambda)}, \label{eq:x_transfer_bar}
\end{equation}

\noindent and from this definition it follows that:
\begin{equation}
	\lbrace \bar{\mathfrak{t}}_S(\lambda), \bar{\mathfrak{t}}_S(\mu) \rbrace^{\-}_S = 0. \nonumber
\end{equation}

\-

Much as in the periodic case, we will consider the generator $ \bar{\mathcal{G}}_S(\lambda) = \ln{\bar{\mathfrak{t}}_S(\lambda)} $, as this will supply us with the known Hamiltonian. To diagonalise the $ T_S^{-1} $, we use:
\begin{equation}
	T_S^{-1}(x, y; -\lambda) = \big( \mathbb{I} + W_S(y; -\lambda) \big) \e{-Z_S(x, y; -\lambda)} \big( \mathbb{I} + W_S(x; -\lambda) \big)^{-1}, \nonumber
\end{equation}

\noindent in place of \eqref{eq:x_WZW}. Consequently, as the highest order term in $ Z_S $ is $ \lambda^{-1} $, the effect of the $ - $ sign outside of the $ Z_S $ and the change in sign of the $ \lambda $ will cancel out, so that the expansion of the exponential term in the limit $ \lambda \to 0^{+} $ is:
\begin{equation}
	\e{-Z_S(x, y; -\lambda)} \to \e{-Z_{11, S}(x, y; -\lambda)} e_{11} + \mathcal{O} (\e{-\lambda^{-1}}). \label{eq:x_OpenLim}
\end{equation}

\noindent Consequently, the expansion of the generator $ \bar{\mathcal{G}}_S $ is:
\begin{equation}
\begin{aligned}
	\bar{\mathcal{G}}_S(\lambda) &= Z_{11, S}(\lambda) - Z_{11, S}(-\lambda) + \ln{\left[ \big( \mathbb{I} + W_S(L; -\lambda) \big)^{-1} K_{+}(\lambda) \big( \mathbb{I} + W_S(L; \lambda) \big) \right]_{11}} \\
	&\qquad+ \ln{\left[ \big( \mathbb{I} + W_S(-L; \lambda) \big)^{-1} K_{-}(\lambda) \big( \mathbb{I} + W_S(-L; -\lambda) \big) \right]_{11}},
\end{aligned} \nonumber
\end{equation}

\noindent where the $ [...]_{ij} $ indicates that we are only considering the $ ij $th component of the matrix inside the brackets. If we expand this expression, the order $ \lambda^0 $ coefficient is constant while the order $ \lambda^1 $ coefficient is:
\begin{equation}
\begin{aligned}
	\bar{\mathcal{G}}_S^{(1)} &= \frac{-1}{2c^3} \int_{-L}^{L} \lb S_{+}' S_{-}' + (S_z')^2 \rb \dd x + \frac{1}{2\alpha_{+} c} \big[ 2\delta_{+} S_z + \beta_{+} S_{+} + \gamma_{+} S_{-} \big]_{x = +L} \\
	&\qquad+ \frac{1}{2\alpha_{-} c} \big[ 2\delta_{-} S_z + \beta_{-} S_{+} + \gamma_{-} S_{-} \big]_{x = -L}.
\end{aligned} \label{eq:x_OpenHam}
\end{equation}

\noindent This can be recognised as $ \mathcal{G}_S^{(1)} $ from \eqref{eq:x_Hams}, up to boundary contributions and an overall factor. As $ \mathcal{G}_S^{(0)} $ was associated to the total momentum of the system, and $ \bar{\mathcal{G}}_S^{(0)} $ is trivial, we can infer that the momentum is no longer conserved when boundary conditions are introduced.

\-

By following an analogous derivation to that of \eqref{eq:x_STS}, we can derive the generator of the $ V $-matrices corresponding to the conserved quantities generated by $ \bar{\mathcal{G}}_S $. There are three cases to consider in this setting \cite{ref:ad_bclax}, corresponding to the $ V $-matrices in the bulk (labelled $ \bar{\mathbb{V}}_{\text{B}} $), and the $ V $-matrices lying at each of the two boundaries (labelled $ \bar{\mathbb{V}}_{\pm} $ for the $ x = \pm L $ boundaries, respectively). The generator of the bulk $ V $-matrices is:
\begin{equation}
\begin{aligned}
	\bar{\mathbb{V}}_{\text{B}, b}(x; \lambda, \mu) &= \bar{\mathfrak{t}}_S^{-1}(\mu) \ptr{a}{K_{+, a} (\mu)T_{S, a}(L, x; \mu) r_{ab}(\mu - \lambda) T_{S, a}(x, -L; \mu) K_{-, a}(\mu) T_{S, a}^{-1}(-\mu) \right. \\
	&\qquad\qquad\quad+ \left. K_{+, a}(\mu) T_{S, a}(\mu) K_{-, a}(\mu) T_{S, a}^{-1}(x, -L; -\mu) r_{ab}(\mu + \lambda) T_{S, a}^{-1}(L, x; -\mu)},
\end{aligned} \label{eq:x_OpenSTS(B)}
\end{equation}

\noindent while the generator of the $ V $-matrices at the positive boundary is:
\begin{equation}
	\bar{\mathbb{V}}_{+, b}(\lambda, \mu) = \bar{\mathfrak{t}}_S^{-1}(\mu) \ptr{a}{K_{-, a}(\mu) T_{S, a}^{-1}(-\mu) \Big( K_{+, a}(\mu) r_{ab}(\mu - \lambda) + r_{ab}(\mu + \lambda) K_{+, a}(\mu) \Big) T_{S, a}(\mu)}, \label{eq:x_OpenSTS(+)}
\end{equation}

\noindent and the generator of the $ V $-matrices at the negative boundary is:
\begin{equation}
	\bar{\mathbb{V}}_{-, b}(\lambda, \mu) = \bar{\mathfrak{t}}_S^{-1}(\mu) \ptr{a}{K_{+, a}(\mu) T_{S, a}(\mu) \Big( r_{ab}(\mu - \lambda) K_{-, a}(\mu) + K_{-, a}(\mu) r_{ab}(\mu + \lambda) \Big) T_{S, a}^{-1}(-\mu)}. \label{eq:x_OpenSTS(-)}
\end{equation}

If we expand these three generators about $ \mu $ as $ \mu \to 0^{+} $, the order $ \mu^0 $ contributions from each generator are trivial, corresponding to $ \bar{\mathcal{G}}_S^{(0)} $ being constant. At order $ \mu^1 $, they are:
\begin{equation}
\begin{aligned}
	\bar{\mathbb{V}}_{\text{B}}^{(1)}(x; \lambda) &= \frac{-1}{2\lambda^2} \mathbb{I} - \frac{1}{2c\lambda^2} S + \frac{1}{2c^3 \lambda} S'S, \\
	\bar{\mathbb{V}}_{\pm}^{(1)}(\lambda) &= \frac{-1}{2\lambda^2} \mathbb{I} - \frac{1}{2c\lambda^2} S \pm \frac{1}{4\alpha_{\pm} c\lambda} \lb \begin{matrix}
		\beta_{\pm} S_{+} - \gamma_{\pm} S_{-} & 2(\delta_{\pm} S_{-} - \beta_{\pm} S_z) \\
		2(\gamma_{\pm} S_z - \delta_{\pm} S_{+}) & \gamma_{\pm} S_{-} - \beta_{\pm} S_{+}
	\end{matrix} \rb.
\end{aligned} \label{eq:VMats_Open}
\end{equation}

\-

In order to extract the boundary conditions from the open Hamiltonian, we simply calculate the equations of motion as usual (through the Poisson brackets and Hamilton's equation), except gathering all of the boundary terms that arise (either from the integration of total derivatives in the bulk Hamiltonian, or from the Poisson bracket of the fields with the boundary Hamiltonians). We then impose the sewing conditions that the equations of motion away from the boundary smoothly transition to those at the boundary, i.e. that $ \lim_{x \to \pm L} \dot{S}_{\sigma}(x) = \dot{S}_{\sigma}(\pm L) $.

Similarly, in order to extract the boundary conditions from the $ V $-matrices, the condition that the equations of motion agree at the boundary manifests as the condition that $ \lim_{x \to \pm L} \bar{\mathbb{V}}_{\text{B}, b} = \bar{\mathbb{V}}_{\pm, b} $. Performing either of these limits yields the same constraint on the boundary constants and the $ S_{\sigma} $ at the boundary \cite{ref:dk_ll}:
\begin{equation}
\begin{aligned}
	\alpha_{\pm} \big[ S_{+} S_{-}' - S_{+}' S_{-} \big]_{x = \pm L} &= \pm c^2 \big[ \beta_{\pm} S_{+} - \gamma_{\pm} S_{-} \big]_{x = \pm L}, \\
	\alpha_{\pm} \big[ S_{+} S_z' - S_{+}' S_z \big]_{x = \pm L} &= \pm c^2 \big[ \delta_{\pm} S_{+} - \gamma_{\pm} S_z \big]_{x = \pm L}, \\
	\alpha_{\pm} \big[ S_{-} S_z' - S_{-}' S_z \big]_{x = \pm L} &= \pm c^2 \big[ \delta_{\pm} S_{-} - \beta_{\pm} S_z \big]_{x = \pm L}.
\end{aligned} \label{eq:x_BC}
\end{equation}

\-

\section{The Dual Model}
\label{sec:Dual}

By considering the equal prominence of the space and time coordinates in the Lagrangian picture of a 1+1 dimensional system, a dual Hamiltonian formulation of the non-linear Schr\"odinger model was constructed in \cite{ref:ck_dual}, which had equal-space Poisson brackets (in place of the equal-time Poisson brackets) and dual integrals of motion that are conserved with respect to space-evolution rather than time-evolution. In this paper we focus on the Lax pair construction rather than the Lagrangian picture emphasised in previous work.

In this Section, we build the dual construction of the isotropic Landau-Lifshitz model in the language of Lax pairs. It follows mostly in parallel with Section \ref{sec:Usual}, with the only divergences being where we emphasise important differences between the two pictures, such as in the limiting procedure of the exponential in the case of open boundary conditions, and where we digress to give an example of how this dual picture can be used to find integrable systems depending non-trivially on additional fields.

The final subsection \ref{ssec:t_Open} considers the introduction of time-like boundary conditions. This idea was introduced in \cite{ref:dfs_nls}, where it was applied to the non-linear Schr\"odinger model.	

\-

\subsection{Poisson Brackets}
\label{ssec:t_PBs}

The first step in this dual construction is defining the equal-space Poisson brackets \eqref{eq:vec_t_PBs} through the use of the $ r $-matrix and an analogue of the linear algebraic relation \eqref{eq:x_LinAlg}. However, as the hierarchy will now describe a series of commuting space flows, the $ S_{\sigma}' $ in the $ V $-matrix \eqref{eq:Lax} will all be derivatives with respect to a specific space-flow, namely the 0th order flow $ x_0 $ (as will be seen later). Consequently, to prevent later confusion, we define these as some new fields, $ \Sigma_{\sigma} $. When we look at the 0th order Hamiltonian or $ V $-matrix (that is, those that provide the original equations of motion \eqref{eq:EoMs}), we will find as part of the space-evolution equations the identification $ \Sigma_{\sigma} = \pd_{x_0} S_{\sigma} $. Otherwise, these $ \Sigma_{\sigma} $ will be treated as entirely independent fields, as can be seen in Subsection \ref{ssec:HigherSys}.

With these new fields, the $ V $-matrix that we consider is:
\begin{equation}
	V = \frac{1}{2\lambda^2} S - \frac{1}{2c^2 \lambda} \Sigma S, \label{eq:VDual}
\end{equation}

\noindent with:
\begin{equation}
	\Sigma = \lb \begin{matrix}
		\Sigma_z & \Sigma_{-} \\
		\Sigma_{+} & -\Sigma_z
	\end{matrix} \rb. \nonumber
\end{equation}

\-

While the Poisson brackets were found from the $ U $- and $ r $-matrices via \eqref{eq:x_LinAlg}, we assume that a similar equation exists for the $ V $-matrices, namely \cite{ref:acdk_dual}:
\begin{equation}
	\lbrace V_a(t_1, \lambda), V_b(t_2, \mu) \rbrace^{\-}_T = [r_{ab}(\lambda - \mu), V_a(t_1, \lambda) + V_b(t_2, \mu)] \,\delta(t_1 - t_2). \label{eq:t_LinAlg}
\end{equation}

\noindent Inserting both the $ V $-matrix and the $ r $-matrix into this expression, we find a collection of Poisson brackets between the various fields:
\begin{equation}
\begin{aligned}
	\lbrace S_{\pm}(t_1), S_z(t_2) \rbrace^{\-}_T &= \lbrace S_{+}(t_1), S_{-}(t_2) \rbrace^{\-}_T = 0, \\
	\lbrace S_{\pm}(t_1), \Sigma_z(t_2) \rbrace^{\-}_T &= \lbrace S_z(t_1), \Sigma_{\pm}(t_2) \rbrace^{\-}_T = S_{\pm} S_z \,\delta(t_1 - t_2), \\
	\lbrace S_z(t_1), \Sigma_z(t_2) \rbrace^{\-}_T &= -S_{+} S_{-} \,\delta(t_1 - t_2), \\
	\lbrace S_{\pm}(t_1), \Sigma_{\pm}(t_2) \rbrace^{\-}_T &= S_{\pm}^2 \,\delta(t_1 - t_2), \\
	\lbrace S_{\pm}(t_1), \Sigma_{\mp}(t_2) \rbrace^{\-}_T &= -(2S_z^2 + S_{+} S_{-}) \,\delta(t_1 - t_2), \\
	\lbrace \Sigma_{\pm}(t_1), \Sigma_z(t_2) \rbrace^{\-}_T &= (S_{\pm} \Sigma_z - \Sigma_{\pm} S_z) \,\delta(t_1 - t_2), \\
	\lbrace \Sigma_{+}(t_1), \Sigma_{-}(t_2) \rbrace^{\-}_T &= (S_{+} \Sigma_{-} - \Sigma_{+} S_{-}) \,\delta(t_1 - t_2).
\end{aligned} \label{eq:t_PBs}
\end{equation}

\noindent As well as sharing the Casimir element $ c^2 = S_z^2 + S_{+} S_{-} $ with the original model, these brackets have an additional commuting quantity:
\begin{equation}
	\tilde{c} = 2S_z \Sigma_z + S_{+} \Sigma_{-} + S_{-} \Sigma_{+}, \label{eq:Casimir2}
\end{equation}

\noindent where, in reference to when $ \Sigma_{\sigma} = \pd_{x_0} S_{\sigma} $ in the HM model, we choose to set $ \tilde{c} = 0 $. Consequently, when the HM model is considered and we can write the $ \Sigma_{\sigma} $ directly as the derivatives of the $ S_{\sigma} $, \eqref{eq:Casimir2} becomes redundant as it is merely the derivative of the original Casimir, \eqref{eq:Casimir}. At any other level of the hierarchy however, we cannot directly relate the $ \Sigma_{\sigma} $ and the $ S_{\sigma} $, so the two Casimirs are distinct.

Introducing the fields $ \Sigma_x $, $ \Sigma_y $, and $ \Sigma_z $ in analogy to $ S_x $, $ S_y $, and $ S_z $, these Poisson brackets can be written more compactly by using the indices $ i, j \in \lbrace x, y, z \rbrace $:
\begin{equation}
\begin{aligned}
	\lbrace S_i(t_1), S_j(t_2) \rbrace^{\-}_T &= 0, \\
	\lbrace S_i(t_1), \Sigma_j(t_2) \rbrace^{\-}_T &= (S_i S_j - c^2 \delta_{ij}) \,\delta(t_1 - t_2), \\
	\lbrace \Sigma_i(t_1), \Sigma_j(t_2) \rbrace^{\-}_T &= (S_i \Sigma_j - S_j \Sigma_i) \,\delta(t_1 - t_2),
\end{aligned} \label{eq:vec_t_PBs}
\end{equation}

\noindent where the two Casimir elements are now:
\begin{equation}
\begin{aligned}
	c^2 &= S_x^2 + S_y^2 + S_z^2, \\
	0 &= S_x \Sigma_x + S_y \Sigma_y + S_z \Sigma_z.
\end{aligned} \label{eq:vec_t_Casimir}
\end{equation}

By defining the quantities:
\begin{equation}
	\psi_1 = S_x^2, \qquad \phi_1 = \frac{1}{2c^2} \lb \frac{\Sigma_z}{S_z} - \frac{\Sigma_x}{S_x} \rb, \qquad\qquad \psi_2 = S_y^2, \qquad \phi_2 = \frac{1}{2c^2} \lb \frac{\Sigma_z}{S_z} - \frac{\Sigma_y}{S_y} \rb, \label{eq:Canon}
\end{equation}

\noindent the above Poisson brackets can be written as a canonical pair (where we use the 2 Casimir elements to discount two of the fields):
\begin{equation}
	\lbrace \psi_1(t_1), \psi_2(t_2) \rbrace^{\-}_T = \lbrace \phi_1(t_1), \phi_2(t_2) \rbrace^{\-}_T = 0, \qquad\qquad \lbrace \psi_i(t_1), \phi_j(t_2) \rbrace^{\-}_T = \delta_{ij} \delta(t_1 - t_2). \label{eq:Canon_PBs}
\end{equation}

\-

\subsection{Periodic Boundary Conditions}
\label{ssec:t_Closed}

In both this section and the next (where open boundary conditions are considered), we consider a system that lies on the interval $ [-\tau, \tau] $, for some $ \tau > 0 $. The periodic boundary conditions in this setting are then $ S_{\sigma}(\tau) = S_{\sigma}(-\tau) $ and $ \Sigma_{\sigma}(\tau) = \Sigma_{\sigma}(-\tau) $.

\-

The construction of the dual model follows in parallel with Section \ref{ssec:x_Closed}. The first object constructed is therefore the equal-space monodromy matrix, $ T_T $, which is a solution to the temporal half of the auxiliary linear problem, \eqref{eq:ALP}, in place of $ \Psi $. This is diagonalised (by analogy to the standard picture discussed in Section \ref{sec:Usual}) through the use of a diagonal matrix $ Z_T $ and an anti-diagonal matrix $ W_T $:
\begin{equation}
\begin{aligned}
	T_T(t_1, t_2; \lambda) &= \text{P}\exp{\int_{t_2}^{t_1} V(\xi) \dd \xi} \\
	&= \big( \mathbb{I} + W_T(t_1; \lambda) \big) \e{Z_T(t_1, t_2; \lambda)} \big( \mathbb{I} + W_T(t_2; \lambda) \big)^{-1}.
\end{aligned} \label{eq:t_Monom}
\end{equation}

\noindent Because we have chosen that the $ V $-matrices satisfy a linear algebraic relation of the form \eqref{eq:t_LinAlg}, the full equal-space monodromy matrix $ T_T(\lambda) = T_T(\tau, -\tau; \lambda) $ will satisfy a quadratic algebraic relation analogous to \eqref{eq:x_QuadAlg}:
\begin{equation}
	\lbrace T_{T, a}(\lambda), T_{T, b}(\mu) \rbrace^{\-}_T = [r_{ab}(\lambda - \mu), T_{T, a}(\lambda) T_{T, b}(\mu)]. \label{eq:t_QuadAlg}
\end{equation}

\-

Taking the trace of the equal-space monodromy matrix we get the equal-space transfer matrix, $ \mathfrak{t}_T $:
\begin{equation}
\begin{aligned}
	\mathfrak{t}_T(\lambda) &= \tr{T_T(\lambda)} \\
	&= \e{Z_{11, T}(\lambda)} + \e{Z_{22, T}(\lambda)},
\end{aligned} \label{eq:t_transfer}
\end{equation}

\noindent which, by virtue of the equal-space monodromy matrix satisfying the quadratic relation \eqref{eq:t_QuadAlg}, Poisson commute for different spectral parameters:
\begin{equation}
	\lbrace \mathfrak{t}_T(\lambda), \mathfrak{t}_T(\mu) \rbrace^{\-}_T = 0. \nonumber
\end{equation}

\noindent Finally, as these two series Poisson commute, so will each pair of the coefficients $ \mathfrak{t}_T^{(k)} $. Therefore, if we take the logarithm of these, $ \mathcal{G}_T(\lambda) = \ln{\mathfrak{t}_T(\lambda)} $, we have that the coefficients in the series expansion of $ \mathcal{G}_T(\lambda) $ Poisson commute with one another:
\begin{equation}
	\lbrace \mathcal{G}_T^{(k)}, \mathcal{G}_T^{(j)} \rbrace^{\-}_T = 0. \label{eq:t_HamsComm}
\end{equation}

\-

As in Section \ref{ssec:x_Closed}, in order to expand $ \mathcal{G}_T $, we need to consider the leading order contribution in each of $ Z_{11, T} $ and $ Z_{22, T} $. Consequently, if we insert the diagonalisation of $ T_T $ into the temporal half of the auxiliary linear problem, \eqref{eq:ALP}, then we find relations for the $ W_T $ and $ Z_T $:
\begin{equation}
\begin{aligned}
	0 &= \dot{W}_T + [W_T, V_D] + W_T V_A W_T - V_A, \\
	\dot{Z}_T &= V_D + V_A W_T,
\end{aligned} \label{eq:t_WZ}
\end{equation}

\noindent where now $ V_D $ and $ V_A $ are the diagonal and anti-diagonal components of the $ V $-matrix, respectively. Expanding $ W_T $ and $ Z_T $ in powers of $ \lambda $ as\footnote{Note that due to the underlying $ V $-matrix having a dependence on $ \lambda^{-2} $ (as compared to the earlier construction where the underlying $ U $-matrix depended only on $ \lambda^{-1} $), the $ Z_T $ series needs to start at $ k = -2 $ instead of $ k = -1 $.}:
\begin{equation}
	W_T(\lambda) = \sum_{k = 0}^{\infty} \lambda^k W_T^{(k)}, \qquad\qquad\qquad Z_T(\lambda) = \sum_{k = -2}^{\infty} \lambda^k Z_T^{(k)}, \nonumber
\end{equation}

\noindent then we can recursively solve \eqref{eq:t_WZ}. Solving the first few orders of these, we find the first three $ Z_T $-matrices to be:
\begin{equation}
\begin{aligned}
	Z_T^{(-2)} &= c\tau \lb \begin{matrix}
		1 & 0 \\
		0 & -1
	\end{matrix} \rb, \qquad\qquad\qquad Z_T^{(-1)} = 0, \\
	Z_T^{(0)} &= \frac{1}{2c} \int_{-\tau}^{\tau} \left[ \dot{S}_z \mathbb{I} + (c - S_z) \lb \begin{matrix}
		\frac{\dot{S}_{-}}{S_{-}} & 0 \\
		0 & -\frac{\dot{S}_{+}}{S_{+}}
	\end{matrix} \rb - \frac{1}{2c^2} (\Sigma_{+} \Sigma_{-} + \Sigma_z^2) \lb \begin{matrix}
		1 & 0 \\
		0 & -1
	\end{matrix} \rb \right] \dd t.
\end{aligned} \label{eq:t_Zs}
\end{equation}

\-

Then, due to the form of the highest order term, the $ \e{Z_{11, T}} $ dominate over the $ \e{Z_{22, T}} $ in \eqref{eq:t_transfer}, so that $ \mathcal{G}_T = Z_{11, T} + ... $. I.e. the first three conserved quantities generated this way will be:
\begin{equation}
\begin{aligned}
	\mathcal{G}_T^{(-2)} &= c\tau, \qquad\qquad\qquad\mathcal{G}_T^{(-1)} = 0, \\
	\mathcal{G}_T^{(0)} &= \frac{1}{2c} \int_{-\tau}^{\tau} \lb \dot{S}_z + (c - S_z) \frac{\dot{S}_{-}}{S_{-}} - \frac{1}{2c^2} (\Sigma_{+} \Sigma_{-} + \Sigma_z^2) \rb \dd t.
\end{aligned} \label{eq:t_Hams}
\end{equation}

\noindent Focussing on the third of these, if we use the periodic boundary conditions to remove any total derivatives and multiply by a factor of $ -2c $, $ \mathcal{G}_T^{(0)} $ reduces to:
\begin{equation}
	H_T = \frac{1}{2} \int_{-L}^{L} \lb \frac{\dot{S}_{+} S_{-} - S_{+} \dot{S}_{-}}{c + S_z} + \frac{1}{c^2} (\Sigma_{+} \Sigma_{-} + \Sigma_z^2) \rb \dd t. \label{eq:t_Ham}
\end{equation}

\noindent This is the equal-space Hamiltonian for the HM model, i.e. the generator of the space-evolution along the space flow $ x_0 $, as can be seen by using $ H_T $ in Hamilton's equation to find the space-evolution equations:
\begin{equation}
	S_{\sigma}' = \lbrace H_T, S_{\sigma} \rbrace^{\-}_T, \qquad\qquad\qquad \Sigma_{\sigma}' = \lbrace H_T, \Sigma_{\sigma} \rbrace^{\-}_T. \nonumber
\end{equation}

\noindent Doing so, the space-evolution equations for $ S_{\sigma} $ simply give the identification $ S_{\sigma}' = \Sigma_{\sigma} $, which is similar to the sine-Gordon model (which has been studied in this description in \cite{ref:c_dual}) and the dual construction of the NLS model \cite{ref:acdk_dual}, while the space-evolution equations for $ \Sigma_{\sigma} $ give:
\begin{equation}
\begin{aligned}
	\Sigma_{\pm}' &= \pm (S_{\pm} \dot{S}_z - \dot{S}_{\pm} S_z) - \frac{1}{c^2} S_{\pm} (\Sigma_{+} \Sigma_{-} + \Sigma_z^2), \\
	\Sigma_z' &= \frac{1}{2} (\dot{S}_{+} S_{-} - S_{+} \dot{S}_{-}) - \frac{1}{c^2} S_z (\Sigma_{+} \Sigma_{-} + \Sigma_z^2),
\end{aligned} \label{eq:Dual_EoMs}
\end{equation}

\noindent which, after substituting in $ S_{\sigma}' = \Sigma_{\sigma} $ can be compactly written as:
\begin{equation}
	\vec{S}'' = \text{i} \vec{S} \times \dot{\vec{S}} - \frac{1}{c^2} \vec{S} |\vec{S}'|^2, \label{eq:Dual_EoMs_vec}
\end{equation}

\noindent and are equivalent to the original equations of motion, \eqref{eq:EoMs}, after replacing $ S_x, S_y $ with $ S_{\pm} = S_x \pm \text{i} S_y $.

\-

Using the equal space Poisson brackets and the tower of equal space conserved quantities, we can generate a whole hierarchy of space-evolution equations associated to distinct systems. Consequently, we will also be interested in generating Lax pairs for each of these systems. By following the derivation of \eqref{eq:x_STS} and \eqref{eq:x_STS_final}, we can derive a generator $ \mathbb{U} $ for the tower of $ U $-matrices that partner with the underlying $ V $-matrix, \eqref{eq:VDual}, which can be generally written as:
\begin{equation}
	\mathbb{U}_b(t; \lambda, \mu) = \mathfrak{t}_T^{-1}(\mu) \ptr{a}{T_{T, a}(\tau, t; \mu) r_{ab}(\mu - \lambda) T_{T, a}(t, -\tau; \mu)}, \label{eq:t_STS}
\end{equation}

\noindent or by using the known results and properties for the $ r $-matrix, as well as the diagonalisation of $ T_T $, this can be reduced to an expression that lies only in one vector space:
\begin{equation}
	\mathbb{U}(t; \lambda, \mu) = \frac{1}{2(\mu - \lambda)} \big( \mathbb{I} + W_T(t; \mu) \big) e_{11} \big( \mathbb{I} + W_T(t; \mu) \big)^{-1}. \label{eq:t_STS_final}
\end{equation}

\noindent When we expand this generator about $ \mu \to 0^{+} $, the first three terms are:
\begin{equation}
\begin{aligned}
	\mathbb{U}^{(0)} &= \frac{-1}{4\lambda} \mathbb{I} - \frac{1}{4c\lambda} S, \\
	\mathbb{U}^{(1)} &= \frac{-1}{4\lambda^2} \mathbb{I} - \frac{1}{4c\lambda^2} S + \frac{1}{4c^3 \lambda} \Sigma S, \\
	\mathbb{U}^{(2)} &= \frac{-1}{4\lambda^3} \mathbb{I} - \frac{1}{4c\lambda^3} S + \frac{1}{4c^3 \lambda^2} \Sigma S + \frac{1}{4c^3 \lambda} \dot{S} S - \frac{1}{8c^5 \lambda} \Sigma^2 S.
\end{aligned} \label{eq:UMats}
\end{equation}

\noindent If we remove the constant factor from the first of these and multiply by a factor of $ -2c $, $ \mathbb{U}^{(0)} $ can be identified with the spatial component of the original Lax pair \eqref{eq:Lax}:
\begin{equation}
	U = -2c (\mathbb{U}^{(0)} + \frac{1}{4\lambda} \mathbb{I}). \nonumber
\end{equation}

\noindent This guarantees that the equations of motion for this model agree with the original equations, \eqref{eq:EoMs}.

\-

\subsection{Higher Order Systems}
\label{ssec:HigherSys}

The identification of the $ \Sigma_{\sigma} $ with the derivatives of the $ S_{\sigma} $ appears as part of the equations of motion for the system at order 0 in the hierarchy (the isotropic Landau-Lifshitz model). If we instead consider a different system, these will not necessarily be the same. To see this, we consider the system at order $ \mu^2 $ in the hierarchy, which has Lax pair $ (U_2, V) $, where we define:
\begin{align}
	U_2 &= -2c (\mathbb{U}^{(2)} + \frac{1}{4\lambda^3} \mathbb{I}) \nonumber \\
	&= \frac{1}{2\lambda^3} S - \frac{1}{2c^2 \lambda^2} \Sigma S - \frac{1}{2c^2 \lambda} \dot{S} S + \frac{1}{4c^4 \lambda} \Sigma^2 S. \label{eq:U2}
\end{align}

\noindent Inserting this Lax pair into the zero-curvature condition, we find the space-evolution equations for this new system. The space-evolution of the three original fields, $ S_{\pm} $ and $ S_z $, are:
\begin{equation}
\begin{aligned}
	S_{+}' &= \frac{1}{c^2} (S_{+} \dot{\Sigma}_z - S_z \dot{\Sigma}_{+}) + \frac{1}{2c^4} (\Sigma_z^2 + \Sigma_{+} \Sigma_{-}) \Sigma_{+}, \\
	S_{-}' &= \frac{1}{c^2} (S_z \dot{\Sigma}_{-} - S_{-} \dot{\Sigma}_z) + \frac{1}{2c^4} (\Sigma_z^2 + \Sigma_{+} \Sigma_{-}) \Sigma_{-}, \\
	S_z' &= \frac{1}{2c^2} (S_{-} \dot{\Sigma}_{+} - S_{+} \dot{\Sigma}_{-}) + \frac{1}{2c^4} (\Sigma_z^2 + \Sigma_{+} \Sigma_{-}) \Sigma_{z},
\end{aligned} \label{eq:U2_EoMs_S}
\end{equation}

\noindent while the space-evolution of the three fields $ \Sigma_{\pm} $ and $ \Sigma_z $ are:
\begin{align}
&\begin{aligned}
	\Sigma_{+}' &= \frac{1}{c^2} (\Sigma_{+} \dot{\Sigma}_z - \dot{\Sigma}_{+} \Sigma_z) + \ddot{S}_{+} + S_{+} \Big( \frac{1}{c^2} \big( (\dot{S}_z)^2 + \dot{S}_{+} \dot{S}_{-} \big) - \frac{1}{2c^6} \big( \Sigma_z^2 + \Sigma_{+} \Sigma_{-} \big)^2 \Big) \\
	&\qquad+ \frac{1}{2c^4} \big( \Sigma_z^2 (\dot{S}_{+} S_z - S_{+} \dot{S}_z) + \Sigma_{+}^2 (\dot{S}_{-} S_z - S_{-} \dot{S}_z) + \Sigma_{+} \Sigma_z (\dot{S}_{+} S_{-} - S_{+} \dot{S}_{-}) \big),
\end{aligned} \nonumber \\
&\begin{aligned}
	\Sigma_{-}' &= \frac{1}{c^2} (\dot{\Sigma}_{-} \Sigma_z - \Sigma_{-} \dot{\Sigma}_z) + \ddot{S}_{-} + S_{-} \Big( \frac{1}{c^2} \big( (\dot{S}_z)^2 + \dot{S}_{+} \dot{S}_{-} \big) - \frac{1}{2c^6} \big( \Sigma_z^2 + \Sigma_{+} \Sigma_{-} \big)^2 \Big) \\
	&\qquad+ \frac{1}{2c^4} \big( \Sigma_z^2 (S_{-} \dot{S}_z - \dot{S}_{-} S_z) + \Sigma_{-}^2 (S_{+} \dot{S}_z - \dot{S}_{+} S_z) + \Sigma_{-} \Sigma_z (\dot{S}_{+} S_{-} - S_{+} \dot{S}_{-}) \big),
\end{aligned} \label{eq:U2_EoMs_Sigma} \\
&\begin{aligned}
	\Sigma_z' &= \frac{1}{2c^2} (\dot{\Sigma}_{+} \Sigma_{-} - \Sigma_{+} \dot{\Sigma}_{-}) + \ddot{S}_{z} + S_{z} \Big( \frac{1}{c^2} \big( (\dot{S}_z)^2 + \dot{S}_{+} \dot{S}_{-} \big) - \frac{1}{2c^6} \big( \Sigma_z^2 + \Sigma_{+} \Sigma_{-} \big)^2 \Big) \\
	&\qquad+ \frac{1}{2c^4} \big( \Sigma_{-} \Sigma_z (S_{+} \dot{S}_z - \dot{S}_{+} S_z) + \Sigma_{+} \Sigma_z (\dot{S}_{-} S_z - S_{-} \dot{S}_z) + \frac{1}{2} (\Sigma_z^2 - \Sigma_{+} \Sigma_{-}) (\dot{S}_{+} S_{-} - S_{+} \dot{S}_{-}) \big).
\end{aligned} \nonumber
\end{align}

\noindent These can be written more compactly in terms of the vectors $ \vec{S} = (S_x, S_y, S_z)^T $ and $ \vec{\Sigma} = (\Sigma_x, \Sigma_y, \Sigma_z)^T $ as:
\begin{equation}
\begin{aligned}
	\vec{S}' &= \frac{\text{i}}{c^2} (\vec{S} \times \dot{\vec{\Sigma}}) + \frac{1}{2c^4} |\vec{\Sigma}|^2 \vec{\Sigma}, \\
	\vec{\Sigma}' &= \frac{\text{i}}{c^2} (\vec{\Sigma} \times \dot{\vec{\Sigma}}) - \frac{\text{i}}{2c^4} |\vec{\Sigma}|^2 (\vec{S} \times \dot{\vec{S}}) + \ddot{\vec{S}} + \vec{S} \Big( \frac{1}{c^2} |\dot{\vec{S}}|^2 - \frac{1}{2c^6} |\vec{\Sigma}|^4 \Big) + \frac{\text{i}}{c^4} \vec{\Sigma} \big( \vec{\Sigma} \cdot (\vec{S} \times \dot{\vec{S}}) \big).
\end{aligned} \label{eq:U2_EoMs_vec}
\end{equation}

\-

\-

When deriving the above Lax pair and resulting equations of motion we started from a $ V $-matrix at order $ \mu^1 $ and found the corresponding $ U $-matrix at order $ \mu^2 $. We could instead, however, start by considering a $ U $-matrix at order $ \mu^2 $ and use that to find the corresponding $ V $-matrix at order $ \mu^1 $.

To find this order $ \mu^2 $ $ U $-matrix, we start from the base system (i.e. the Lax pair consisting of the $ U $- and $ V $-matrices appearing at order $ \mu^0 $, see \eqref{eq:VMats} and \eqref{eq:UMats}):
\begin{equation}
	U = V = \frac{1}{2\lambda} S. \label{eq:Base_Lax}
\end{equation}

\noindent The equations of motion for this system are simply $ \dot{S}_{\sigma} = S'_{\sigma} $. Then, the first three terms in the hierarchy of $ U $-matrices constructed from the $ V $-matrix are:
\begin{equation}
\begin{aligned}
	\mathbb{U}^{(0)} &= \frac{-1}{4\lambda} \mathbb{I} - \frac{1}{4c\lambda} S, \\
	\mathbb{U}^{(1)} &= \frac{-1}{4\lambda^2} \mathbb{I} - \frac{1}{4c\lambda^2} S + \frac{1}{4c^3 \lambda} \dot{S} S, \\
	\mathbb{U}^{(2)} &= \frac{-1}{4\lambda^3} \mathbb{I} - \frac{1}{4c\lambda^3} S + \frac{1}{4c^3 \lambda^2} \dot{S} S - \frac{1}{4c^3\lambda} \ddot{S} - \frac{1}{8c^5\lambda} (\dot{S})^2 S,
\end{aligned} \label{eq:Base_UMats}
\end{equation}

\noindent which should be compared with \eqref{eq:VMats}. Before we can construct the space-like (standard) hierarchy for the $ U $-matrix found from $ \mathbb{U}^{(2)} $ we need to define the fields $ P_{\sigma} = \pd_{t_0} S_{\sigma} $ and $ \mathbb{P}_{\sigma} = \pd_{t_0}^2 S_{\sigma} $ (in analogy to how we defined the field $ \Sigma_{\sigma} = \pd_{x_0} S_{\sigma} $), so that the $ U $-matrix is:
\begin{equation}
	U = \frac{1}{2\lambda^3} S - \frac{1}{2c^2\lambda^2} PS + \frac{1}{2c^2\lambda} \mathbb{P} + \frac{3}{4c^4\lambda} P^2 S, \label{eq:U2_comm}
\end{equation}

\noindent with:
\begin{equation}
	P = \lb \begin{matrix}
		P_z & P_{-} \\
		P_{+} & -P_z
	\end{matrix} \rb, \qquad\qquad\qquad \mathbb{P} = \lb \begin{matrix}
		\mathbb{P}_z & \mathbb{P}_{-} \\
		\mathbb{P}_{+} & -\mathbb{P}_z
	\end{matrix} \rb. \nonumber
\end{equation}

\noindent This is the $ U $-matrix appearing at order $ \mu^2 $ that we consider in place of \eqref{eq:U2}. Constructing the space-like hierarchy from this, the $ V $-matrix appearing at order $ \mu^1 $ is (after removing the constant factor and scaling by $ -2c $):
\begin{equation}
	V = \frac{1}{2\lambda^2} S - \frac{1}{2c^2\lambda} PS. \label{eq:V1_comm}
\end{equation}

This Lax pair would appear to describe a system of equations different to \eqref{eq:U2_EoMs_vec}, due to containing a total of nine fields, $ S_{\sigma} $, $ P_{\sigma} $, and $ \mathbb{P}_{\sigma} $. When these matrices are inserted into the zero-curvature condition, however, one of these sets of fields is redundant and $ \mathbb{P} $ can be written in terms of $ S $ and $ P $ as:
\begin{equation}
	\mathbb{P} = S\dot{S} - \frac{1}{c^2} P^2 S. \nonumber
\end{equation}

\noindent The combination of this identification and the remaining equations of motion can then be recognised as the equations \eqref{eq:U2_EoMs_vec}. Consequently, traversing the early ($ n < 3 $) part of these dual hierarchies is commutative for this model. It remains to be seen if any higher order parts of the dual hierarchies commute, however, there is no \textit{a priori} justification for the commutativity and an investigation into this is left for future study.

\-

\subsection{Open Boundary Conditions}
\label{ssec:t_Open}

Finally, we consider the effect of introducing reflective boundary conditions to the time-axis. This idea was introduced in \cite{ref:dfs_nls}, where it was applied to the NLS model. Due to the $ r $-matrix structure for the dual model, \eqref{eq:t_LinAlg}, being identical to the $ r $-matrix structure of the original model, \eqref{eq:x_LinAlg}, we introduce boundary conditions in an identical manner. That is, we start by choosing a pair of matrices, $ K_{\pm} $, that satisfy \eqref{eq:cRA}. Specifically, we use the same $ K $-matrices as in the original picture, \eqref{eq:KMat}:
\begin{equation}
	K_{\pm}(\lambda) = \alpha_{\pm} \mathbb{I} + \lambda \lb \begin{matrix}
		\delta_{\pm} & \beta_{\pm} \\
		\gamma_{\pm} & -\delta_{\pm}
	\end{matrix} \rb, \nonumber
\end{equation}

\noindent where the constants $ \alpha_{\pm} $, $ \beta_{\pm} $, $ \gamma_{\pm} $, and $ \delta_{\pm} $ could in general depend on the evolution parameter, $ x $, but we choose them to be constant for simplicity. We introduce these $ K $-matrices into the generator of the quantities conserved with respect to space as \cite{ref:s_qbcs, ref:s_bcs, ref:dfs_nls}:
\begin{equation}
	\bar{\mathfrak{t}}_T(\lambda) = \tr{K_{+}(\lambda) T_T(\tau, -\tau; \lambda) K_{-}(\lambda) T_T^{-1}(\tau, -\tau; -\lambda)}, \label{eq:t_transfer_bar}
\end{equation}

\noindent from which we can use the quadratic relation \eqref{eq:t_QuadAlg} and the defining relation for the $ K $-matrices, \eqref{eq:cRA}, to derive the time-like equivalent of \eqref{eq:x_transfer_bar}, which tells us that the $ \bar{\mathfrak{t}}_T $ Poisson commute for different spectral parameters. Again, we are actually interested in the coefficients in the expansion of $ \bar{\mathcal{G}}_T(\lambda) = \ln{\bar{\mathfrak{t}}_T(\lambda)} $, which will also Poisson commute with one another:
\begin{equation}
	\lbrace \bar{\mathcal{G}}_T^{(k)}, \bar{\mathcal{G}}_T^{(j)} \rbrace^{\-}_T = 0. \label{eq:t_OpenHamsComm}
\end{equation}

In order to evaluate the series expansion of $ \bar{\mathcal{G}}_T(\lambda) $, as well as diagonalising $ T_T $ through \eqref{eq:t_Monom}, we need to also diagonalise $ T_T^{-1} $ through:
\begin{equation}
	T_T^{-1}(t_1, t_2; -\lambda) = \big( \mathbb{I} + W_T(t_2; -\lambda) \big) \e{-Z_T(t_1, t_2; -\lambda)} \big( \mathbb{I} + W_T(t_1; -\lambda) \big)^{-1}. \nonumber
\end{equation}

\noindent An important point here is that when we take the limit as $ \lambda \to 0^{+} $ of the exponentiated term, due to the $ - $ sign in front of the $ Z_T $ and the highest order term being $ (-\lambda)^2 = \lambda^2 $, the expansion of the exponential as $ \lambda \to 0^{+} $ will instead be:
\begin{equation}
	\e{-Z_T(t_1, t_2; -\lambda)} \to \e{-Z_{22, T}(t_1, t_2; -\lambda)} e_{22} + \mathcal{O}(\e{-\lambda^{-2}}). \nonumber
\end{equation}

\noindent Consequently, when the diagonalisations are inserted into the generator $ \bar{\mathcal{G}}_T $, we have (where we suppress the parameters by defining $ \hat{f} = f(-\lambda) $ and $ W_{\pm, T} = W_T(\pm \tau) $):
\begin{equation}
	\bar{\mathcal{G}}_T(\lambda) = \ln{\e{Z_{11, T} - \hat{Z}_{22, T}} \tr{K_{+} \big( \mathbb{I} + W_{+, T} \big) e_{11} \big( \mathbb{I} + W_{-, T} \big)^{-1} K_{-} \big( \mathbb{I} + \hat{W}_{-, T} \big) e_{22} \big( \mathbb{I} + \hat{W}_{+, T} \big)^{-1}}}, \nonumber
\end{equation}

\noindent which can be separated into the bulk contribution and the two boundary contributions:
\begin{equation}
	\bar{\mathcal{G}}_T(\lambda) = Z_{11, T}(\lambda) - Z_{22, T}(-\lambda) + \ln{\mathbb{W}_{+}(\lambda)} + \ln{\mathbb{W}_{-}(\lambda)}, \label{eq:t_OpenHamGen}
\end{equation}

\noindent where we define:
\begin{equation}
\begin{aligned}
	\mathbb{W}_{+}(\lambda) &= \Big[ \big( \mathbb{I} + W_T(\tau; -\lambda) \big)^{-1} K_{+}(\lambda) \big( \mathbb{I} + W_T(\tau; \lambda) \big) \Big]_{21}, \\
	\mathbb{W}_{-}(\lambda) &= \Big[ \big( \mathbb{I} + W_T(-\tau; \lambda) \big)^{-1} K_{-}(\lambda) \big( \mathbb{I} + W_T(-\tau; -\lambda) \big) \Big]_{12}.
\end{aligned} \label{eq:Wbb}
\end{equation}

\-

Due to the logarithmic dependence of $ \bar{\mathcal{G}}_T $ on $ \mathbb{W}_{\pm} $, the lowest order contribution of the boundary terms to the generator $ \bar{\mathcal{G}}_T $ will appear at order $ \lambda^0 $. Specifically, this lowest order contribution will be:
\begin{equation}
	\mathbb{W}_{\pm}^{(1)} = \frac{1}{2c} \bigg( \frac{\pm 2\alpha_{\pm}}{c} \Big( \frac{S_{\pm} \Sigma_z}{S_z + c} - \Sigma_{\pm} \Big) - 2\delta_{\pm} S_{\pm} - \beta_{\pm} \frac{S_{+} S_{\pm}}{S_z \pm c} - \gamma_{\pm} \frac{S_{-} S_{\pm}}{S_z \mp c} \bigg), \label{eq:Wbb_1}
\end{equation}

\noindent so that the first three terms in the expansion of $ \bar{\mathcal{G}}_T $ are:
\begin{equation}
\begin{aligned}
	\bar{\mathcal{G}}_T^{(-2)} &= 2c\tau, \qquad\qquad\qquad \bar{\mathcal{G}}_T^{(-1)} = 0, \\
	\bar{\mathcal{G}}_T^{(0)} &= \frac{1}{2c} \int_{-\tau}^{\tau} \lb \frac{S_{+} \dot{S}_{-} - \dot{S}_{+} S_{-}}{c + S_z} - \frac{1}{c^2} (\Sigma_{+} \Sigma_{-} + \Sigma_z^2) \rb \dd t + \ln{\mathbb{W}_{+}^{(1)}} + \ln{\mathbb{W}_{-}^{(1)}}.
\end{aligned} \label{eq:t_OpenGs}
\end{equation}

\noindent Multiplying $ \bar{\mathcal{G}}_T^{(0)} $ by the factor $ -c $ gives the Hamiltonian with open boundary conditions:
\begin{equation}
	\bar{H}_T = \int_{-\tau}^{\tau} \lb \frac{1}{2c^2} (\Sigma_{+} \Sigma_{-} + \Sigma_z^2) + \frac{\dot{S}_{+} S_{-} - S_{+} \dot{S}_{-}}{2(c + S_z)} \rb \dd t - c\ln{\mathbb{W}_{+}^{(1)}} - c\ln{\mathbb{W}_{-}^{(1)}}. \label{eq:t_OpenHam}
\end{equation}

\-

Away from the boundaries, the Poisson brackets of $ \bar{H}_T $ with each of the six fields returns the space-evolution equations, \eqref{eq:Dual_EoMs}. At the boundaries, however, when the space-evolution is derived the condition that the fields at the boundary still satisfy the usual space-evolution equations imposes extra conditions on the fields $ S_{\sigma} $ and $ \Sigma_{\sigma} $, as well as the $ \alpha_{\pm} $, $ \beta_{\pm} $, $ \gamma_{\pm} $, and $ \delta_{\pm} $. The requirement that $ \lim_{t \to \pm \tau} S_{\sigma}' = S_{\sigma}'(\pm \tau) $ restricts us to the case $ \alpha_{\pm} = 0 $. If we combine this with the requirement that $ \lim_{t \to \pm \tau} \Sigma_{\sigma}' = \Sigma_{\sigma}'(\pm \tau) $, then we find the time-like boundary conditions for the HM model:
\begin{equation}
	0 = \alpha_{\pm}, \qquad\qquad\qquad 0 = \beta_{\pm} S_{+} + \gamma_{\pm} S_{-} + 2\delta_{\pm} S_z. \label{eq:t_BC}
\end{equation}

\-

We can also find a generator for the $ U $-matrices both in the bulk and at the boundaries. The generator for the bulk $ U $-matrices will be \cite{ref:dfs_nls}:
\begin{equation}
\begin{aligned}
	\bar{\mathbb{U}}_{\text{B}, b}(t; \lambda, \mu) &= \bar{\mathfrak{t}}_T^{-1}(\mu) \ptr{a}{K_{+, a} (\mu) T_{T, a}(\tau, t; \mu) r_{ab}(\mu - \lambda) T_{T, a}(t, -\tau; \mu) K_{-, a}(\mu) T_{T, a}^{-1}(-\mu) \right. \\
	&\qquad\qquad\quad+ \left. K_{+, a}(\mu) T_{T, a}(\mu) K_{-, a}(\mu) T_{T, a}^{-1}(t, -\tau; -\mu) r_{ab}(\mu + \lambda) T_{T, a}^{-1}(\tau, t; -\mu)},
\end{aligned} \label{eq:t_OpenSTS(B)}
\end{equation}

\noindent and, being mindful of the different limit for the $ T_T^{-1}(-\mu) $, this can be reduced to:
\begin{equation}
	\bar{\mathbb{U}}_{\text{B}}(t; \lambda, \mu) = \mathbb{U}(t; \lambda, \mu) + \frac{1}{2(\mu + \lambda)} \big( \mathbb{I} + W_T(t; -\mu) \big) e_{22} \big( \mathbb{I} + W_T(t; -\mu) \big)^{-1}, \label{eq:t_OpenSTS_final}
\end{equation}

\noindent where $ \mathbb{U}(t; \lambda, \mu) $ is the generator of the $ U $-matrices with periodic boundary conditions. Unlike in the original case, where the second term differed from the first only by the sign of the $ \mu $, here it differs both by the sign of the $ \mu $ and in that the matrix $ e_{11} $ has become $ e_{22} $. The lowest order term in the expansion of this appears as the coefficient of $ \mu^0 $, and is:
\begin{equation}
	\mathbb{U}_{\text{B}}^{(0)} = \frac{-1}{2c\lambda} \lb \begin{matrix}
		S_z & S_{-} \\
		S_{+} & -S_z
	\end{matrix} \rb = 2\mathbb{U}^{(0)}, \label{eq:U(B)}
\end{equation} 

\noindent where $ \mathbb{U}^{(0)} $ is the $ U $-matrix appearing at lowest order in the periodic case. The boundary $ U $-matrices are found by considering the generators:
\begin{equation}
\begin{aligned}
	\bar{\mathbb{U}}_{+, b}(\lambda, \mu) &= \bar{\mathfrak{t}}_T^{-1}(\mu) \ptr{a}{K_{-, a}(\mu) T_{T, a}^{-1}(-\mu) \Big( K_{+, a}(\mu) r_{ab}(\mu - \lambda) + r_{ab}(\mu + \lambda) K_{+, a}(\mu) \Big) T_{T, a}(\mu)}, \\
	\bar{\mathbb{U}}_{-, b}(\lambda, \mu) &= \bar{\mathfrak{t}}_T^{-1}(\mu) \ptr{a}{K_{+, a}(\mu) T_{T, a}(\mu) \Big( r_{ab}(\mu - \lambda) K_{-, a}(\mu) + K_{-, a}(\mu) r_{ab}(\mu + \lambda) \Big) T_{T, a}^{-1}(-\mu)},
\end{aligned} \label{eq:t_OpenSTS(pm)}
\end{equation}

\noindent which can be simplified to:
\begin{equation}
\begin{aligned}
	\bar{\mathbb{U}}_{+, b}(\lambda, \mu) &= \frac{1}{2\mathbb{W}_{+}(\mu)} \lb \frac{1}{\mu - \lambda} \big( \mathbb{I} + W_T(\tau; \mu) \big) e_{12} \big( \mathbb{I} + W_T(\tau; -\mu) \big)^{-1} K_{+}(\mu) \right. \\
	&\qquad\qquad\quad+\left. \frac{1}{\mu + \lambda} K_{+}(\mu) \big( \mathbb{I} + W_T(\tau; \mu) \big) e_{12} \big( \mathbb{I} + W_T(\tau; -\mu) \big)^{-1} \rb,
\end{aligned} \label{eq:t_OpenSTS(+)}
\end{equation}

\noindent and:
\begin{equation}
\begin{aligned}
	\bar{\mathbb{U}}_{-, b}(\lambda, \mu) &= \frac{1}{2\mathbb{W}_{-}(\mu)} \lb \frac{1}{\mu - \lambda} K_{-}(\mu) \big( \mathbb{I} + W_T(-\tau; -\mu) \big) e_{21} \big( \mathbb{I} + W_T(-\tau; \mu)) \big)^{-1} \right. \\
	&\qquad\qquad\quad+\left. \frac{1}{\mu + \lambda} \big( \mathbb{I} + W_T(-\tau; -\mu) \big) e_{21} \big( \mathbb{I} + W_T(-\tau; \mu) \big)^{-1} K_{-}(\mu) \rb.
\end{aligned} \label{eq:t_OpenSTS(-)}
\end{equation}

\noindent The first non-trivial term in the expansion of each of these appears at order $ \mu^0 $. For the $ t = +\tau $ boundary, this is:
\begin{equation}
\begin{aligned}
	\mathbb{U}_{+}^{(0)} &= \frac{1}{2c (c + S_z) \mathbb{W}_{+}^{(1)}} \left[ \frac{\alpha_{+}}{\lambda^2} \lb \begin{matrix}
		S_{+} (c + S_z) & -(c + S_z)^2 \\
		S_{+}^2 & -S_{+} (c + S_z)
	\end{matrix} \rb \right. \\
	&\qquad\left.- \frac{1}{2\lambda} \lb \begin{matrix}
		-\beta_{+} S_{+}^2 - \gamma_{+} (c + S_z)^2 & 2(c + S_z) \big( \delta_{+} (c + S_z) + \beta_{+} S_{+} \big) \\
		2S_{+} \big( \delta_{+} S_{+} - \gamma_{+} (c + S_z) \big) & \beta_{+} S_{+}^2 + \gamma_{+} (c + S_z)^2
	\end{matrix} \rb \right],
\end{aligned} \label{eq:U(+)}
\end{equation}

\noindent while at the $ t = -\tau $ boundary, the $ U $-matrix is:
\begin{equation}
\begin{aligned}
	\mathbb{U}_{-}^{(0)} &= \frac{1}{2c (c + S_z) \mathbb{W}_{-}^{(1)}} \left[ \frac{\alpha_{-}}{\lambda^2} \lb \begin{matrix}
		S_{-} (c + S_z) & S_{-}^2 \\
		-(c + S_z)^2 & -S_{-} (c + S_z)
	\end{matrix} \rb \right. \\
	&\qquad\left.- \frac{1}{2\lambda} \lb \begin{matrix}
		\beta_{-} (c + S_z)^2 + \gamma_{-} S_{-}^2 & -2S_{-} \big( \delta_{-} S_{-} - \beta_{-} (c + S_z) \big) \\
		-2(c + S_z) \big( \delta_{-} (c + S_z) + \gamma_{-} S_{-} \big) & -\beta_{-} (c + S_z)^2 - \gamma_{-} S_{-}^2
	\end{matrix} \rb \right].
\end{aligned} \label{eq:U(-)}
\end{equation}

\noindent Requiring that $ \lim_{t \to \pm \tau} \mathbb{U}_{\text{B}}^{(0)} = \mathbb{U}_{\pm}^{(0)} $ gives rise to both the condition that $ \alpha_{\pm} = 0 $ (from the order $ \lambda^{-2} $ terms) and that $ \beta_{\pm} S_{+} + \gamma_{\pm} S_{-} + 2\delta_{\pm} S_z = 0 $, which agrees with the boundary conditions found from the Hamiltonian approach, \eqref{eq:t_BC}.

\-

By comparing the time-like boundary conditions, \eqref{eq:t_BC}, with the space-like boundary conditions, \eqref{eq:x_BC}, we can see that there is no evident connection between the two. This asymmetry is rooted in the fundamentally different dependence of the fields on the space and time coordinates, as can be seen by comparing the forms of the equations of motion in \eqref{eq:EoMs_vec} and \eqref{eq:Dual_EoMs_vec}.

\-

\section{Summary}
\label{sec:Outro}

The main result of this paper, derived in Section \ref{sec:Dual}, is the dual construction of the isotropic Landau-Lifshitz model, where space-evolution equations, spatially conserved quantities, and equal-space Poisson brackets are obtained. This was done by following the usual procedure for deriving Poisson brackets and conserved quantities for a system that is integrable via the existence of a Lax pair and $ r $-matrix, except with the roles of the space and time variables switched. A consequence of this equal-space construction is the existence of a hierarchy of dual integrable systems, each of which has an infinite tower of conserved quantities, \eqref{eq:t_Hams}, and a Lax pair representation, \eqref{eq:UMats}. Then, through the combination of the usual equal-time hierarchy and this dual equal-space hierarchy, an infinite ``lattice" of integrable models can be built (it is important to note here that this ``lattice" is not commutative \textit{a priori}, although it has been observed to commute for $ n, m < 3 $).

By considering a higher order system in the dual hierarchy of the isotropic Landau-Lifshitz model, \eqref{eq:U2_EoMs_vec}, we have connected the 3-field HM model (with 1 Casimir element) with a novel 6-field model (which has 2 Casimir elements). As this system appears in the hierarchy of the HM model, it is likely to have a solitonic solution similar to that of the HM model, which would be discoverable through the use of the inverse scattering tools, or through a Darboux-B\"acklund/Dressing approach. The investigation of such a soliton could provide interesting insights into the dual construction, if not the original model itself, but we leave this for future consideration.

\-

We have also studied the introduction of reflective boundary conditions to the time-axis in Section \ref{ssec:t_Open}, in the vein of \cite{ref:dfs_nls}. While seemingly unphysical, such boundary conditions could have applications as a particular type of initial condition for the system, where the time coordinate is considered on the half-line, $ [0, \infty) $, instead. Thus, the boundary conditions discussed above would appear as a particular set of initial conditions that settle into (in the case of a soliton reflecting boundary) a 2-soliton solution. Potential applications and consequences of this however are left for later investigation.

\-

Finally, we close by repeating that, due to the $ U $- and $ V $-matrices sharing the same $ r $-matrix, the space and time coordinates in this construction are fully interchangeable. This means that all of the results described here will still hold when the space and time coordinates are switched, so that switching the space derivatives and time derivatives in \eqref{eq:U2_EoMs_vec} describes the time-evolution of an integrable system:
\begin{equation}
\begin{aligned}
	\dot{\vec{S}} &= \frac{\text{i}}{c^2} (\vec{S} \times \vec{\Sigma}') + \frac{1}{2c^4} |\vec{\Sigma}|^2 \vec{\Sigma}, \\
	\dot{\vec{\Sigma}} &= \frac{\text{i}}{c^2} (\vec{\Sigma} \times \vec{\Sigma}') - \frac{\text{i}}{2c^4} |\vec{\Sigma}|^2 (\vec{S} \times \vec{S}') + \vec{S}'' + \vec{S} \Big( \frac{1}{c^2} |\vec{S}'|^2 - \frac{1}{2c^6} |\vec{\Sigma}|^4 \Big) + \frac{\text{i}}{c^4} \vec{\Sigma} \big( \vec{\Sigma} \cdot (\vec{S} \times \vec{S}') \big),
\end{aligned} \label{eq:tEvo_Higher}
\end{equation}

\noindent and the results of Section \ref{ssec:t_Open} can be viewed instead as a description of (space-like) open boundary conditions for the time-evolution equations:
\begin{equation}
	\ddot{\vec{S}} = \text{i} (\vec{S} \times \vec{S}') - \frac{1}{c^2} \vec{S} |\dot{\vec{S}}|^2. \label{eq:tEvo_Dual}
\end{equation}

\-

This dual construction has now been applied to the isotropic Landau-Lifshitz model, the non-linear Schr\"odinger model (originally in scalar \cite{ref:ck_dual} case and later extended to the vector \cite{ref:zlg_dual} case) and its associated hierarchy (including, for example, the complex modified KdV equation) in \cite{ref:acdk_dual}, and the sine-Gordon model in \cite{ref:c_dual}. All of these models can be found as special limits of the anisotropic Landau-Lifshitz model \cite{ref:s_linalg} and its hierarchy. Consequently, it would be expected that the fully anisotropic Landau-Lifshitz model also admits a space-time duality of this type, however, an investigation into this is left for future work.

\-

\subsubsection*{Acknowledgements}

The author would like to thank the EPSRC funding council for a PhD studentship, and his PhD supervisor Anastasia Doikou for feedback and encouragement. He would also like to thank Calum Ross and Lukas M\"uller for proofreading and comments, as well as the reviewer for useful feedback.

\-

\raggedright

\-


\begin{thebibliography}{99}

\bibitem{ref:acdk_dual}
	J. Avan, V. Caudrelier, A. Doikou, A. Kundu, ``Lagrangian and Hamiltonian structures in an integrable hierarchy and space-time duality", \textit{Nucl. Phys.} \textbf{B902} (2016), 415-39, \href{https://doi.org/10.1016/j.nuclphysb.2015.11.024}{doi:10.1016/j.nuclphysb.2015.11.024}

\bibitem{ref:ck_dual}
	V. Caudrelier, A. Kundu, ``A multisymplectic approach to defects in integrable classical field theory", \textit{J. High Energ. Phys.} \textbf{02} (2015) 88, \href{http://dx.doi.org/10.1007/JHEP02(2015)088}{doi:10.1007/JHEP02(2015)088}
	
\bibitem{ref:l_lax}
    P. D. Lax, ``Integrals of nonlinear equations of evolution and solitary waves", \textit{Comm. Pure. Appl. Math.} \textbf{21} (1968) 467-90, \href{https://doi.org/10.1002/cpa.3160210503}{doi:10.1002/cpa.3160210503}

\bibitem{ref:akns_lax}
	M. J. Ablowitz, D. J. Kaup, A. C. Newell, H. Segur, ``The Inverse Scattering Transform‐Fourier Analysis for Nonlinear Problems", \textit{Stud. Appl. Math.} \textbf{53} (1974), 249-315, \href{https://doi.org/10.1002/sapm1974534249}{doi:10.1002/sapm1974534249}
	
\bibitem{ref:ac_dual}
	J. Avan, V. Caudrelier, ``On the origin of dual Lax pairs and their r-matrix structure", \textit{J. Geom. Phys.} \textbf{120} (2017), 106-28, \href{https://doi.org/10.1016/j.geomphys.2017.05.010}{doi:10.1016/j.geomphys.2017.05.010}

\bibitem{ref:l_hm}
	M. Lakshmanan, ``Continuum spin system as an exactly solvable dynamical system", \textit{Phys. Lett.} \textbf{61A} (1977) 53-4, \href{https://doi.org/10.1016/0375-9601(77)90262-6}{doi:10.1016/0375-9601(77)90262-6}

\bibitem{ref:t_hm}
	L. A. Takhtajan, ``Integration of the continuous Heisenberg spin chain through the inverse scattering method", \textit{Phys. Lett.} \textbf{64A} (1977) 235-7, \href{https://doi.org/10.1016/0375-9601(77)90727-7}{doi:10.1016/0375-9601(77)90727-7}

\bibitem{ref:s_linalg}
	E. K. Sklyanin, ``On complete integrability of the Landau-Lifshitz equation". Preprint LOMI E-3-79, Leningrad 1979

\bibitem{ref:ft_book}
	L. D. Faddeev, L. A. Takhtajan, \textit{Hamiltonian Methods in the Theory of Solitons}, Springer-Verlag 1987, \href{https://doi.org/10.1007/978-3-540-69969-9}{doi:10.1007/978-3-540-69969-9}

\bibitem{ref:dfs_nls}
	A. Doikou, I. Findlay, S. Sklaveniti, ``Time-like boundary conditions in the NLS model", \textit{Nucl. Phys.} \textbf{B941} (2019) 361-75, \href{https://doi.org/10.1016/j.nuclphysb.2019.02.022}{doi:10.1016/j.nuclphysb.2019.02.022}
	
\bibitem{ref:c_dual}
	V. Caudrelier, ``Multisymplectic approach to integrable defects in the sine-Gordon model", \textit{J. Phys.} \textbf{A48} (2015) 195203, \href{https://doi.org/10.1088/1751-8113/48/19/195203}{doi:10.1088/1751-8113/48/19/195203}

\bibitem{ref:s_qbcs}
	E. K. Sklyanin, ``Boundary conditions for integrable quantum systems", \textit{J. Phys.} \textbf{A21} (1988), 2375-89, \href{https://doi.org/10.1088/0305-4470/21/10/015}{doi:10.1088/0305-4470/21/10/015}

\bibitem{ref:s_bcs}
	E. K. Sklyanin, ``Boundary conditions for integrable equations", \textit{Funct. Anal. Its. Appl.} \textbf{21} (1987), 164-6, \href{https://doi.org/10.1007/BF01078038}{doi:10.1007/BF01078038}

\bibitem{ref:dk_ll}
	A. Doikou, N. Karaiskos, ``Generalized Landau–Lifshitz models on the interval", \textit{Nucl. Phys.} \textbf{B853} (2011), 436-60, \href{https://doi.org/10.1016/j.nuclphysb.2011.08.001}{doi:10.1016/j.nuclphysb.2011.08.001}

\bibitem{ref:ads_hmlim}
	J. Avan, A. Doikou, K. Sfetsos, ``Systematic classical continuum limits of integrable spin chains and emerging novel dualities", \textit{Nucl. Phys.} \textbf{B840} (2010), 469-90, \href{https://doi.org/10.1016/j.nuclphysb.2010.07.014}{doi:10.1016/j.nuclphysb.2010.07.014}
	
\bibitem{ref:f_book}
    E. Fradkin, \textit{Field Theories of Condensed Matter Physics}, Frontiers in Physics \textbf{82}, Addison-Wesley (1991), \href{https://doi.org/10.1017/CBO9781139015509}{doi:10.1017/CBO9781139015509}

\bibitem{ref:dlsvv_hm}
	F. Demontis, S. Lombardo, M. Sommacal, C. van der Mee, F. Vargiu, ``Effective generation of closed-form soliton solutions of the continuous classical Heisenberg ferromagnet equation", \textit{Commun. Nonlinear Sci. Numer. Simulat.} \textbf{64} (2018) 35–65, \href{https://doi.org/10.1016/j.cnsns.2018.03.020}{doi:10.1016/j.cnsns.2018.03.020}

\bibitem{ref:msp+_magnets}
	S.M. Mohseni, S.R. Sani, J. Persson, \textit{et al.}, ``Spin Torque–Generated Magnetic Droplet Solitons", \textit{Science} \textbf{339} (2013) 1295-8, \href{https://doi.org/10.1126/science.1230155}{doi:10.1126/science.1230155}

\bibitem{ref:ls_magnets}
	J. W. Lau, J. M. Shaw, ``Magnetic nanostructures for advanced technologies: fabrication, metrology and challenges", \textit{J. Phys. D: Appl. Phys.} \textbf{44} (2011) 303001, \href{https://doi.org/10.1088/0022-3727/44/30/303001}{doi:10.1088/0022-3727/44/30/303001}

\bibitem{ref:sts_vgen}
	M. A. Semenov-Tian-Shansky, ``What is a classical r-matrix?", \textit{Funct. Anal. Appl.} \textbf{17} (1983), 259-72, \href{https://doi.org/10.1007/BF01076717}{doi:10.1007/BF01076717}

\bibitem{ref:stf_frt}
	E. K. Sklyanin, L. A. Takhtajan, L. D. Faddeev, ``Quantum inverse problem method. I", \textit{Theoret. and Math. Phys.} \textbf{40}:2 (1979), 688-706, \href{https://doi.org/10.1007/BF01018718}{doi:10.1007/BF01018718}

\bibitem{ref:frt_frt}
	N. Yu. Reshetikhin, L. A. Takhtajan, L. D. Faddeev, ``Quantization of Lie Groups and Lie Algebras", \textit{Leningrad Math. J.}, \textbf{1}:1 (1990), 193-225

\bibitem{ref:dg_kmat}
	H. J. de Vega, A. Gonz{\'a}lez-Ruiz, ``Boundary K-matrices for the XYZ, XXZ and XXX spin chains", \textit{J. Phys.} \textbf{A27} (1994), 6129-38, \href{https://doi.org/10.1088/0305-4470/27/18/021}{doi:10.1088/0305-4470/27/18/021}

\bibitem{ref:ad_bclax}
	J. Avan, A. Doikou, ``Integrable boundary conditions and modified Lax equations", \textit{Nucl. Phys.} \textbf{B800} (2008), 591-612, \href{https://doi.org/10.1016/j.nuclphysb.2008.04.004}{doi:10.1016/j.nuclphysb.2008.04.004}
	
\bibitem{ref:zlg_dual}
	R.-G. Zhou, P.-Y. Li, Y. Gao, ``Equal-Time and Equal-Space Poisson Brackets of the N-Component Coupled NLS Equation", \textit{Commun. Theor. Phys.} \textbf{67} (2017) 347-9, \href{https://doi.org/10.1088/0253-6102/67/4/347}{doi:10.1088/0253-6102/67/4/347}

\end{thebibliography}
\end{document}